 \documentclass[pmlr,twocolumn,10pt]{jmlr} 

\usepackage{makecell} 

\usepackage{booktabs}
\usepackage{siunitx}

\usepackage[switch]{lineno}

\newcommand{\equal}[1]{{\hypersetup{linkcolor=black}\thanks{#1}}}

\theorembodyfont{\upshape}
\theoremheaderfont{\scshape}
\theorempostheader{:}
\theoremsep{\newline}

\jmlrvolume{LEAVE UNSET}
\jmlryear{2024}
\jmlrsubmitted{LEAVE UNSET}
\jmlrpublished{LEAVE UNSET}
\jmlrworkshop{Machine Learning for Health (ML4H) 2024} 

 \title[Diffusion-Based Segmentation of Lumbar MRI of Lower Back Pain Patients]{Diffusion-Based Semantic Segmentation of Lumbar Spine MRI Scans of Lower Back Pain Patients}

\author{%
\Name{Maria Monzon}\equal{These authors contributed equally} \Email{mmonzon@ethz.ch}\\
\addr ETH Zurich, Switzerland
\AND
\Name{Thomas Iff}\footnotemark[1] \Email{thomas.iff@inf.ethz.ch}\\
\addr ETH Zurich, Switzerland
\AND
\Name{Ender Konukoglu}
\Email{ender.konukoglu@vision.ee.ethz.ch}\\
\addr ETH Zurich, Switzerland
\AND
\Name{Catherine R. Jutzeler} \Email{catherine.jutzeler@hest.ethz.ch}\\
\addr ETH Zurich, Switzerland
}

\begin{document}

\maketitle

\begin{abstract}
This study introduces a diffusion-based framework for robust and accurate segmenton of vertebrae, intervertebral discs (IVDs), and spinal canal from Magnetic Resonance Imaging~(MRI) scans of patients with low back pain (LBP), regardless of whether the scans are T1w or T2-weighted.
The results showed that SpineSegDiff achieved comparable outperformed non-diffusion state-of-the-art models in the identification of degenerated IVDs. 
Our findings highlight the potential of diffusion models to improve LBP diagnosis and management through precise spine MRI analysis.
\end{abstract}


\paragraph*{Data and Code Availability}

This work uses publicly available SPIDER dataset~\citep{od-SpiderGraaf2023} for training and evaluation. It includes T1w and T2-weighted MRI scans of the lumbar spine from 218 subjects with low back pain. The SpineSegDiff model code, along with training and evaluation scripts, and reproducibility instructions, is available at \url{https://github.com/BMDS-ETH/SpineSegDiff}.

\paragraph*{Institutional Review Board (IRB)}

This research study retrospectively analyzed open access human subject data , exempt from ethical approval according to the open access license of ~\citep{od-SpiderGraaf2023}.

\section{Introduction}
\label{sec:intro}
Low Back Pain~(LBP) is one of the leading causes of disability \citep{Dionne2006-bpp}, affecting more than 600 million of the global population at some point in their lifetime \citep{Hartvigsen2018-mf}. Diagnosing LBP is particularly challenging due to the various pathophysiological mechanisms involved~\citep{Fourney2011ChronicApproach}, including social, genetic, biophysical and psychological factors. 
While Magnetic Resonance Imaging~(MRI) serves as a crucial diagnostic tool, both manual and automated interpretation face significant challenges due toltifaceted nature of LBP. 
Manual evaluation is challenging due to  inter-rater variability and the high labor demand while automated segmentation must overcome three key anatomical challenges: the high intraclass similarity between vertebrae \citep{Wang2022-SpinsegBC, Sekuboyina2021-verse}, substantial  anatomical variability caused by degenerative pathologies such as disc herniation, spinal stenosis, and vertebral fractures \citep{od-MRSegChallengePang2018,od-SpiderGraaf2023}. 
Motivated by inherent ability of diffusion models to model complex and noisy data distributions ~\citep{LiOnModels}, this study presents the following contributions: (i) 
the development of SpineSegDiff, a 2D diffusion-based segmentation model specifically  for semantic segmentation of strucures handling pathological variability present in MRI scans of LBP patients, with an emphasis on managing both T1 and T2-weighted MRI scans within a unified model; and (ii) the implementation of a pre-segmentation strategy that takes advantage accelerated the diffusion models training while mantaining its advantages.

\begin{figure*}[t]
  \centering
  \includegraphics[width=0.65\textwidth]{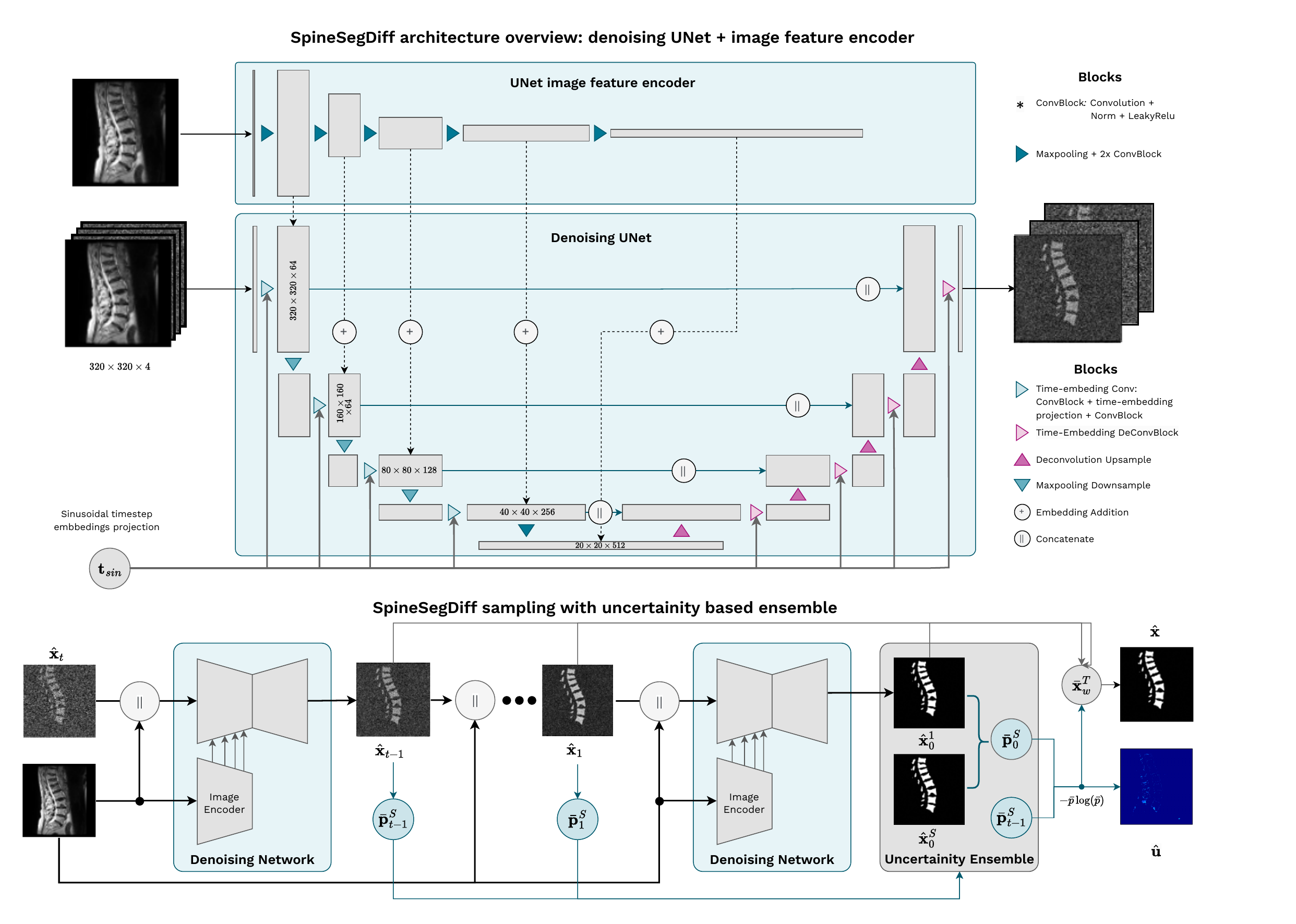}
\caption{ SpineSegDiff architecture overview: the 2D MRI scan ($\mathbf{y}$) is concatenated with the partially noised mask to generate the segmentation image $\mathbf{x}_{T}$.
The architecture is composed by a multi-scale image encoder and a UNet-based model for denoising at each diffusion.
The final mask is inferred using a uncertainty-step ensemble that creates uncertainty maps across multiple samples and time steps.}\label{fig:spisegdiff}
\label{fig:spinsegdiff-overview}
\end{figure*} 

\section{Related Work}
Recently, Denoising diffusion probabilistic models DDPMs)~\citep{Ho2020DenoisingModels}, showed promising results in medical image analysis~\citet{Kazerouni2023DiffusionSurvey,chung2022-diffres} due to their ability to effectively capture the underlying data distributions~\citep{DhariwalDiffusionSynthesis} and handle noise and variability in medical images ~\citep{LiOnModels}.  
Typically used for image generation, can be adapted for segmentation by modeling conditional distributions~$p(\mathbf{x}|\mathbf{y})$, treating the mask as a generated sample $\mathbf{x}$ with the image $\mathbf{y}$ as a condition ~\citep{Wolleb2021DiffusionEnsembles}. 

Several studies have explored DDPM for medical image segmentation \citep{Liu2024diffSurvey,Xing2023Diff-UNet:Segmentation,Wolleb2021DiffusionEnsembles,kim2023-diffadvrep,DS-Wu2022}, highlighting two main diffusion-based methods. 
The Implicit Image Segmentation Diffusion Model (IISDM) \citep{Wolleb2021DiffusionEnsembles}  uses noise prediction to iteratively denoise the segmentation mask . In contrast, DiffUnet directly directly infers the final segmentation mask $\hat{\mathbf{x}}_0$ from a partially noised input $\mathbf{x}_t$.

\begin{figure*}[t]
\floatconts
  {fig:results-statistics}
  {\caption{Qualitative and quantitative analysis of SpineSegDiff performance.}}
  {%
    \subfigure[Qualitative evaluation of segmentation for the diffusion models and baseline on three anatomical structures: spinal canal (blue), vertebrae (green), and intervertebral discs (red)]{\label{fig:results}%
      \includegraphics[width=0.38\textwidth]{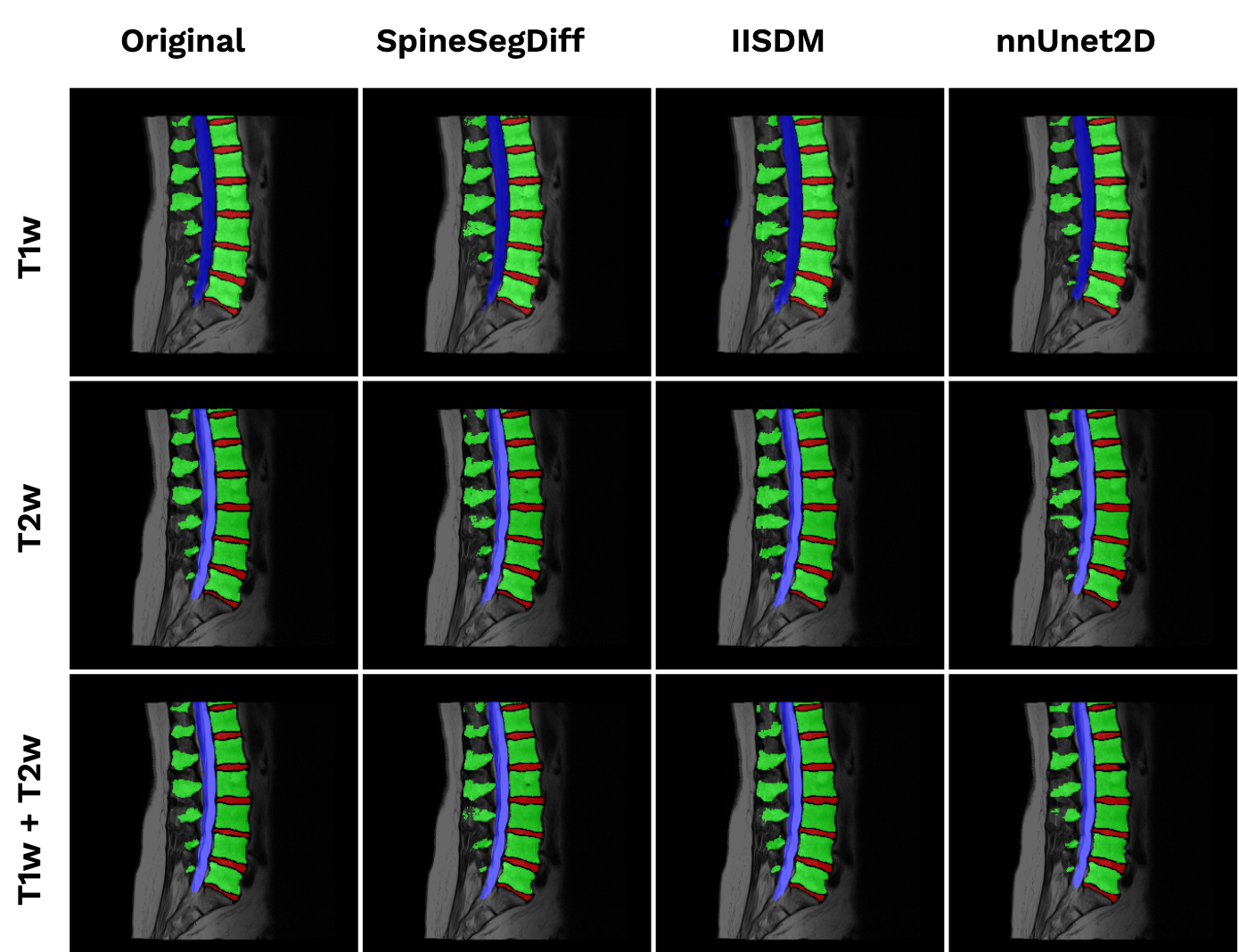}%
    }%
    \qquad
    \subfigure[Quantitative comparison of segmentation performance for  spinal structures: spinal canal (SC), vertebrae, and intervertebral discs (IVD) for SpineSegDiff, IISDM, and nnU-Net models  across T1-weighted (T1w), T2-weighted (T2w), and combined (T1w + T2w) MRI]{%
      \includegraphics[width=0.55\textwidth]{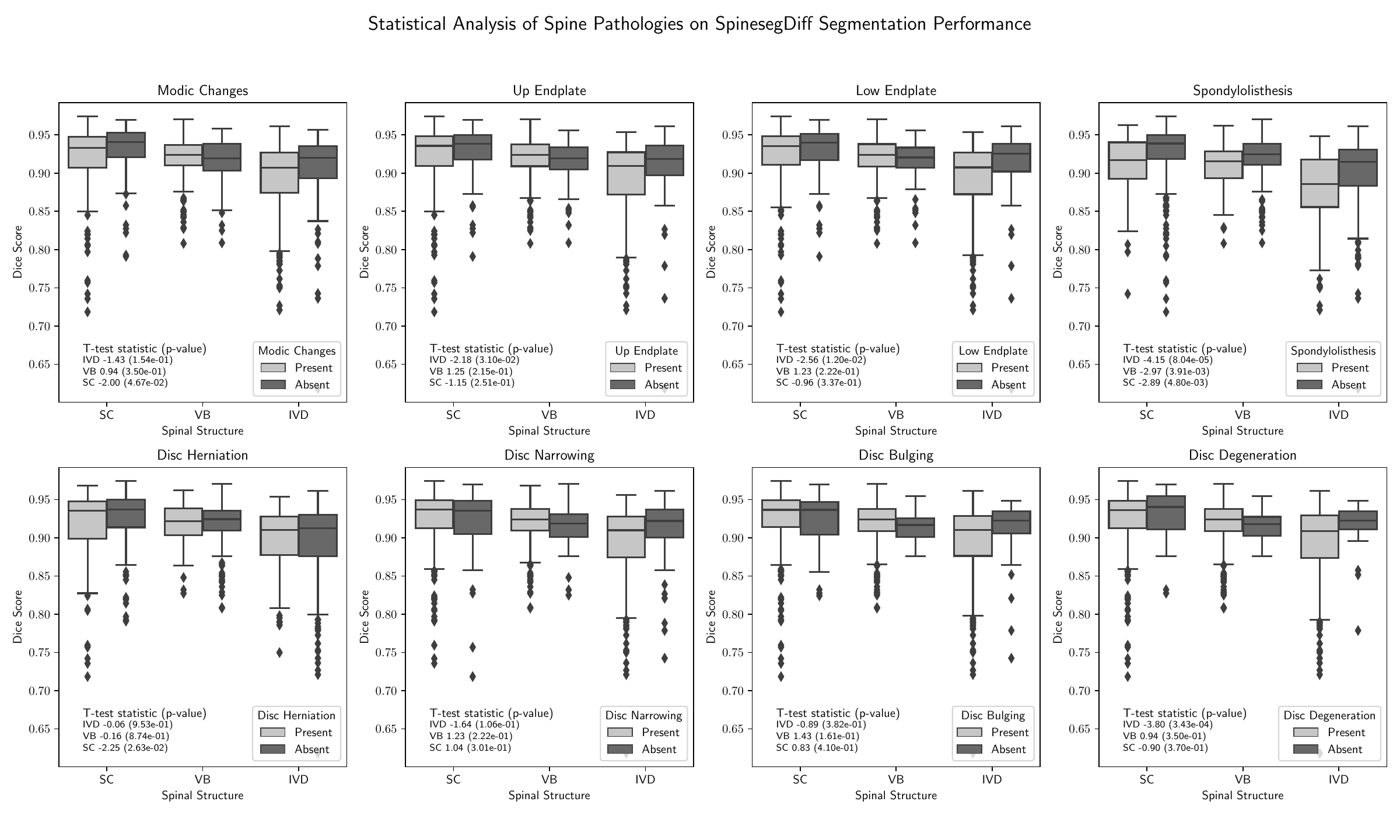}%
    }
  }
\end{figure*}

\section{Methods: SpineSegDiff  Model}
This research presents an innovative two-dimensional diffusion-based framework (Fig.\ref{fig:spinsegdiff-overview})  aimed at achieving semantic segmentation of the central slice in lumbar spine MRI scans for the diagnosis of LBP. In contrast to traditional diffusion models that learn to denoise patterns \citep{Wolleb2021DiffusionEnsembles}, SpineSegDiff directly generates the output mask $\mathbf{x}_0$ .  
SpineSegDiff architecture (appendix \ref{apd:A}) captures multiscale features from the input image. The Denoising UNet,  adds the extract meaningful features from additional encoder with time-extracted embeddings to  enriching the model with supplementary information during training \citep{Xing2023Diff-UNet:Segmentation}. 
SpineSegDiff was trained  using a composite loss function that combines MSE denoising loss for reconstruction with Dice Loss and Binary Cross-Entropy Loss. This approach effectively penalizes discrepancies between the predicted segmentation mask $\hat{\mathbf{x}}_f$ and the ground truth mask $\mathbf{x}_0$. 
The SpineSegDiff inference process employs a step uncertainty ensemble strategy \citep{Xing2023Diff-UNet:Segmentation} to create the final segmentation mask $\hat{\mathbf{x}}_f$ by leveraging the stochastic nature of diffusion models, detailed in Appendix \ref{apd:A}

\subsection{Pre-segmentation training with nnUnet}

SpineSegDiff training is significantly accelerated through the implementation of a pre-segmentation strategy \citep{Guo2022AcceleratingSegmentation}. 
The complete system is composed of nnU-Net followed by a SpineSegDiff architecture.
The initial segmentation $\hat{\mathbf{x}}_\text{pre}$ is predicted with the pre-trained baseline nnU-net model \citep{nnUnet-Isensee2018}.
The diffusion process involves a short diffusion training of SpineSegDiff model, which takes pre-segmentation mask $\hat{\mathbf{x}}_\text{pre}$  partially noised  with a cosine noise scheduler as input and learns to denoise it to recover the original segmentation mask $\mathbf{x}_0$. 

\section{Experimental Results}

\subsection{Dataset}

The investigation used sagittal MRI of the lumbar spine from a cohort of 218 patients (63\% female) within the SPIDER~\citep{od-SpiderGraaf2023} data set,  with pathological conditions such as spondylolisthesis, disc herniation, and modic changes (appendix \ref{apd:B}). 
The MRI preprocessing pipeline included image normalization by 98th percentile and scaling to 0-255 pixel range. Scans were then realigned to the RAS+ coordinate system for consistent orientation. Finally, images were resampled to a uniform spatial resolution of 1mm and resized to 320x320 pixels.

\subsection{Statistical Evaluation of Pathologies}

\begin{table*}[htbp]
\small
\floatconts
  {tab:results-constrast}%
  {\caption{
Quantitative comparison of segmentation performance for  spinal structures: spinal canal (SC), vertebrae, and intervertebral discs (IVD) for SpineSegDiff, IISDM, and nnU-Net models and  across T1-weighted (T1w), T2-weighted (T2w), and combined T1w + T2w MRI modalities.}}%
{
\begin{tabular}{lccccc}

\toprule
\textbf{Model}  &  \textbf{Modality} & \textbf{Spinal Canal} & \textbf{Vertebrae} & \textbf{IVD}& \textbf{mDICE} \\
\midrule
SpineSegDiff &  T1w& 0.93 $\pm$ 0.04 & 0.91 $\pm$ 0.03 & 0.89 $\pm$ 0.05 & 0.908 \\
SpineSegDiff & T2w& \textbf{0.93 $\pm$ 0.04 }& 0.92 $\pm$ 0.04 & \textbf{0.90 $\pm$ 0.04 }& \textbf{0.917} \\
SpineSegDiff &  T1w + T2w& 0.92 $\pm$ 0.04 & \textbf{0.92 $\pm$ 0.02} & 0.90 $\pm$ 0.05 & 0.913\\
\midrule
IISDM &  T1w & 0.87 $\pm$ 0.10 & 0.91 $\pm$ 0.04 & 0.89 $\pm$ 0.05 & 0.890 \\
IISDM & T2w & 0.86 $\pm$ 0.12 & 0.91 $\pm$ 0.04 & 0.89 $\pm$ 0.05 & 0.887\\
IISDM & T1w + T2w & 0.90 $\pm$ 0.03 & 0.92 $\pm$ 0.05 &0.89 $\pm$ 0.04 & 0.903\\
\midrule
nnU-Net & T1w & 0.91 $\pm$ 0.02 & 0.91 $\pm$ 0.03 & 0.84 $\pm$ 0.06 & 0.887 \\
nnU-Net & T2w & 0.91 $\pm$ 0.03 & 0.92 $\pm$ 0.03 & 0.85 $\pm$ 0.06  & 0.893\\
nnU-Net & T1w + T2w & 0.91 $\pm$ 0.03 & 0.92 $\pm$ 0.03 & 0.84 $\pm$ 0.05 & 0.890\\
\bottomrule
\end{tabular}
}
\end{table*}

We performed statistical analysis to assess the impact of different pathologies on segmentation performance across various spine structuresof SpineSegDiff model trained for both T1w+T2w .  
Pathologies such as Modic changes, disc herniation, and spondylolisthesis, disc narrowing, overall disc degeneration evaluated through Pfirrman grading, were considered due to their potential impact on model accuracy. The pathology distribution of the study cohort is detailed in the appendix \ref{apd:B}. 
Figure \ref{fig:results-statistics} depicts box plots of Dice scores between patients with and without these conditions and t-test results that highlight the relationship between these pathologies and model performance. To address the issue of multiple comparisons, we applied the Benjamini-Hochberg p-values correction at $\alpha = 0.05$. 

The figure indicates that pathologies like spondylolisthesis and disc narrowing significantly impact segmentation. Upper endplate changes affected IVD segmentation (p=0.0310), while lower endplate changes impacted both IVD (p=0.0120) and SC (p=0.0337). Spondylolisthesis had widespread effects on SC (p=0.0048), VB (p=0.0039), and IVD (p$<$0.0001) segmentation scores. Disc herniation only significantly affected SC segmentation (p=0.0263), and disc degeneration significantly affected IVD segmentation (p = 0.0003).

\subsection{Evaluating Diffusion Models for MRI Contrast-Independent Segmentation}

\begin{table*}[htbp]
\small
\centering
\floatconts
{tab:presegmentation}{
\caption{ Segmentation performance of pre-segmentation stategy for diffusion timesteps ($T$)}}
{
\begin{tabular}{lcccccc}
\textbf{} & \textbf{$T=0$} & \textbf{$T=30$} & \textbf{$T=100$} & \textbf{$T=300$} & \textbf{$T=500$} & \textbf{$T=1000$} \\
\midrule
\textbf{Spinal Canal} & 0.91 $\pm$ 0.03 & \textbf{0.92 $\pm$ 0.05} & 0.92 $\pm$ 0.06 & 0.92 $\pm$ 0.06 & 0.92 $\pm$ 0.06 & 0.92 $\pm$ 0.07 \\
\textbf{Vertebrae} & 0.92 $\pm$ 0.03 & \textbf{0.92 $\pm$ 0.04} & 0.91 $\pm$ 0.04 & 0.91 $\pm$ 0.04 & 0.91 $\pm$ 0.04 & 0.91 $\pm$ 0.03 \\
\textbf{IVD} & 0.84 $\pm$ 0.05 & \textbf{0.89 $\pm$ 0.05} & 0.89 $\pm$ 0.06 & \textbf{0.89 $\pm$ 0.05} & 0.89 $\pm$ 0.06 & \textbf{0.89 $\pm$ 0.05} \\
\bottomrule
\end{tabular}
}
\end{table*}
The baseline experiment evaluated diffusion models' efficacy in segmenting MRI scans across different contrasts (T1w, T2w, and combined T1w+T2w) without contrast-specific customization. 
To benchmark our model, we also trained the nnU-Net~\citep{Isensee2020NnU-Net:Segmentation} model as a baseline.

Model performance for the segmentation of lumbar structures  was assessed using Dice scores with 5-fold cross-validation, ensuring patient-wise split consistency. 18 oblique MRI scans were excluded from the evaluation but retained for training.
The results summarized in Table \ref{tab:results-constrast}  showed that SpineSegDiff slightly outperformed the benchmarked for all MRI modalities, particularily in the segmentation of the intervertebral disc.

\subsection{Pre-segmentation Time Diffusion Steps} 

We conducted an ablation study  to determine the optimal number of timesteps $t$ that balance computational efficiency and segmentation accuracy, To evaluate the effectiveness of the pre-segmentation strategy. 
Various time-step configurations were tested, and the results were compared to a baseline model using 1000 steps starting from the noised pre-segmentation, summarized in Table \ref{tab:presegmentation}.The baseline model, as outlined in $T=0$, designates the non-diffusion segmentation as a prior. 
The ablation study revealed that the pre-segmentation  strategy significantly reduced the number of time steps needed while maintaining the  segmentation performance.

\section{Discussion and Conclusion}
Our findings demonstrate the potential of diffusion models,
 for accurate and efficient lumbar spine MRI segmentation. 
Particularly SpineSegDiff excells  in IVD disc delineation,  crucial for LBP diagnosis and treatment planning as disc degeneration is a common pain cause.  The statistical analysis reveals that certain degenerative pathologies, particularly spondylolisthesis and disc narrowing, can substantially reduce the accuracy of SpineSegDiff, which exhibit the highest t-statistics and the lowest p-values. 

By leveraging the initial segmentation produced by nnUNet, the study of diffusion time steps $(T)$ needed (Table \ref{tab:presegmentation}) reveals the the pre-segmentation strategy effectively balances accuracy and computational efficiency, making SpineSegDiff more practical for clinical use. Nonetheless,  it is important to acknowledge the limitations of our study and the challenges that remain for clinical translation.
The computational requirements of diffusion models, even with the pre-segmentation strategy, may still pose barriers to widespread adoption, particularly in resource-limited settings.  
To fully realize the potential of SpineSegDiff, future work should focus on two key areas.
First, efforts should be made to further optimize the model's computational efficiency, making it  suitable for clinical implementation. 
Second, despite the multicenter nature of the dataset, the model should be validated on larger and more diverse datasets to ensure its generalizability between different patient populations and imaging protocols.


In conclusion, the work introduces a novel diffusion-based approach for lumbar spine MRI segmentation, achieving state-of-the-art performance in identifying degenerated intervertebral discs.
By implementing a pre-segmentation strategy, the model maintains high accuracy while reducing computational requirements. Despite the current computational challenges, the model's unique ability to quantify segmentation uncertainties offers a promising pathway for more precise detection of low back pain-related pathologies. 
Future research should concentrate on enhancing the computational efficiency and assessing the robustness of the model's applicability across various patient demographics and imaging methodologies.

\acks{This project was supported by grant \# 380 of the Strategic Focus Area \textit{Personalized Health and Related Technologies (PHRT)} of the ETH Domain (Swiss Federal Institutes of Technology). 
The funders did not specify the study design, data collection, analysis, or the decision to publish and preparation of the manuscript.
The manuscript preparation benefited from AI-assisted writing tools including Perplexity AI, Copilot and Writefull. These tools were used for language refinement, clarity enhancement, and writing style improvements. All scientific content, methodology, analysis, and conclusions remain the original work of the authors. The authors reviewed, edited, and take full responsibility for this publication's content after using the tool/service.
}

\bibliography{references}

\newpage

\appendix

\section{SpineSegDiff}\label{apd:A}

\subsection{Architecure Details}

SpineSegDiff employs a  dual-encoder architecture optimized for lumbar spine MRI segmentation. The model integrates a U-shaped backbone, adapted from the optimal nnUNet configuration, with an additional dedicated image encoder \cite{Xing2023Diff-UNet:Segmentation}. 
The supplementary encoder comprises multiple convolutional layers with a feature hierarchy of [64,64,128,256,512,64] and LeakyReLU activations, designed to capture multi-scale anatomical features from the input MRI scans. The architecture's denoising component combines these extracted features with time-embedded information during the diffusion process, enhancing the model's ability to reconstruct anatomical structures progressively. 

\subsection{ Uncertainty Based Inference}

The SpineSegDiff inference process employs a step uncertainty ensemble strategy adapted from \citep{Xing2023Diff-UNet:Segmentation} to create the final segmentation mask $\hat{\mathbf{x}}_f$ by leveraging the stochastic nature of diffusion models and generating $S$ intermediate samples. 

The mean probability $\bar{\mathbf{p}}_t$ for each timestep $t$ is computed as the average of the probability of every sample $s$,
\begin{equation}
 \bar{\mathbf{p}}_t= \frac{1}{S} \sum_{s=1}^{S} \mathbf{p}_s
\end{equation}

where $\mathbf{p}_s$ represents the softmax probability map for each sample $s$ at each diffusion timestep \(t\) during DDIM sampling. 
The uncertainty $\hat{\mathbf{u}}_t$ at the time step $t$ is calculated as the entropy of the average probability distribution, which is the negative log-likelihood of the mean probability distribution across all samples $S$:

\begin{equation}
\hat{\mathbf{u}}_t = -   \bar{\mathbf{p}}_t \cdot  \log \left(  \bar{\mathbf{p}}_t\right)
\end{equation}

The final prediction is calculated as the weighted sum of the samples across $T_s$ last timesteps by its uncertainty $\hat{\mathbf{u}}_t$ scaled by the time. 

\begin{equation}
\hat{\mathbf{x}}_f = \sum_{t=1}^{T_s} \left(e^{\sigma\left(\frac{t}{T_S}\right)} \times (1  - \hat{\mathbf{u}}_t ) \right) \times  \bar{\mathbf{p}}_t 
\end{equation}
where 
$e^{\sigma\left(\frac{t}{T_s}\right)}$  term represents a time-based scaling factor, where $\sigma$ is the sigmoid function. It ensures that predictions from different timesteps are appropriately weighted, with the weight increasing as the timestep progresses.

\section{Experiments Details}\label{apd:B}

\subsection{Dataset: Degenerative Pathologies}
The dataset comprises a multicenter collection of sagittal lumbar MRI obtained from four different hospitals in the Netherlands,
Ground truth labels for semantic segmentation were created by combining vertebrae annotations (starting from L5) and one-hot-encoded into three structures: spinal canal (SC), vertebral bodies (VB), and IVD. 
The incidence of present spinal degenerative pathologies was determined if they manifested at any vertebral level and is summarized in the following table.
\vspace{-0.5 cm}
\begin{table}[thb]
\caption{Overview of degenerative pathology's presence in the SPIDER dataset}
\centering
\label{tab:pathologies}
\begin{tabular}{lr}
\textbf{Pathology} & \textbf{Patients (\%)} \\
\toprule
Spondylolisthesis & 42 (19.27\%) \\
Disc Herniation & 72 (33.03\%)  \\
Modic Changes & 149 (68.34\%) \\
Endplate Changes & 177 (81.19\%) \\
Disc Narrowing & 193 (88.53\%) \\
Disc Bulging & 200 (91.74\%)\\
\end{tabular}
\end{table}

\subsection{Implementation Details} 

The models were trained on a high-performance cluster using RTX 4090 GPUs for 2D and v100 GPUs for 3D models. The models were implemented with Pytorch and MONAI \citep{JorgeCardoso2022MONAI:Healthcare} frameworks. 
The 2D models were trained and evaluated only on the central slice of the data, whereas the 3D models were trained and evaluated on the entire volume. 
The optimal epochs for diffusion models was determined by the segmentation precision \citep{Bertels2019-DICE} in the first-fold validation set, where 2500 epochs were used for SpineSegDiff training. 
The diffusion models training time steps were set to $T = 1000$ with a linear variance noise schedule from $\beta_1 = 10^{-4}$ to 0.02 
The rest of training hyperparameters for all the compared modes are summarized in Appendix \ref{apd:A}.

\subsubsection{SpineSegDiff}  
The SpineSegDiff model is trained using a composite loss function that combines Mean Squared Error (MSE), Dice Loss, and Binary Cross-Entropy (BCE) Loss. The total loss is formulated as:\newline
$
L_{total} =  L_{MSE} +L_{Dice} + L_{BCE}
$
where each terms are can be decomposed as 
$L_{MSE} = \frac{1}{N} \sum_{i=1}^N (\hat{x}_i - x_i)^2$, 
$L_{Dice} = 1 - \frac{2|\hat{X} \cap X|}{|\hat{X}| + |X|}$, 
$L_{BCE} = -\frac{1}{N} \sum_{i=1}^N [x_i \log(\hat{x}_i) + (1-x_i) \log(1-\hat{x}_i)]$. 
This loss optimizes the model for pixel accuracy (MSE), segmentation quality (Dice), and probabilistic output (BCE). The training hyperparameters are summarized in the table below:

\begin{table}[h]
  \caption{Training hyperparameters for SpineSegDiff}
  \centering
  \small
  \begin{tabular}{lccc}
    \toprule
    \textbf{Parameter} & \textbf{T1w, T2w, T1w+T2w}\\ 
    \midrule
    \textbf{Image Size} & 320x320 \\
     \textbf{Epochs} & 2500 \\
    \textbf{Batch} & 4  \\
    \textbf{Optimizer} & AdamW  \\
    \textbf{Learning Rate} & 0.0001\\
  \textbf{Training Loss} & MSE +  Dice +  Cross Entropy \\
    \bottomrule
  \end{tabular}
\end{table}

\subsubsection{nnUnet Baseline}
The nnU-Net model  \citep{Isensee2020NnU-Net:Segmentation}  is trained using a highly automated and adaptable framework designed for semantic segmentation tasks which informs the configuration of multiple U-Net architectures tailored to the dataset's specific characteristics. 
The model training involves a multi-step process that includes preprocessing, model configuration, training. nnU-Net employs a five-fold cross-validation strategy to ensure robust performance evaluation. The training utilizes various configurations, such as 2D, 3D full resolution, to optimize segmentation performance across different data modalities. The hyperparameters that were used in the training are summarzed in the following tables: 
\begin{table}[h]
\caption{Training Hyperparameters for nnUnet 2D}\centering\small\begin{tabular}{lccc}\toprule\textbf{Parameter} & \textbf{T1w} & \textbf{T2w} & \textbf{T1w+T2w} 
\\ \midrule
Patch Size & 256x64 & 256x64 & 256x64 \\
Epochs & 250 & 250 & 250 \\
Batch & 197 & 197 & 197 \\
Optimizer & SGD & SGD & SGD \\
Learning Rate & 0.01 & 0.01 & 0.01 \\
Training Loss & Dice & Dice & Dice \\ \bottomrule\end{tabular}
\end{table}
\vspace{-0.5 cm}
\begin{table}[h]\caption{Training Hyperparameters for nnUnet 3D}\centering\small\begin{tabular}{lccc}\toprule\textbf{Parameter} & \textbf{T1w} & \textbf{T2w} & \textbf{T1w+T2w} \\ \midrule
Patch Size & \makecell{ 56x\\ 224x192} & \makecell{ 56x\\ 224x192} & \makecell{ 56x\\ 224x192}\\
Epochs & 250 & 250 & 250 \\
Batch & 2 & 2 & 2 \\
Optimizer & SGD & SGD & SGD \\
Learning Rate & 0.01 & 0.01 & 0.01 \\
Training Loss & Dice & Dice & Dice \\ \bottomrule\end{tabular}\end{table}

\subsection{Implicit Image Segmentation Diffusion Model (IISMD)}
IISMD ~\citep{Wolleb2021DiffusionEnsembles} follows DDPM training, adding Gaussian noise $\boldsymbol{\epsilon}_t \sim \mathcal{N}(\mathbf{0}, \mathbf{I})$ to the segmentation mask $\mathbf{x}_0$ at each timestep $t \in \{1, \ldots, T\}$ using a linear noise scheduler $\{\alpha_t \in (0, 1)\}_{t=1}^T$. For denoising, U-Net architecture $f_{\boldsymbol{\theta}}$ estimates noise $\boldsymbol{\epsilon}_t = f_{\boldsymbol{\theta}}(\mathbf{x}_t, \mathbf{y}, t)$ at each timestep, concatenated with MRI images $\mathbf{y}$, used to guide the generation of the segmentation mask.  The parameters $\boldsymbol{\theta}$ are optimized by minimizing the Mean Squared Error (MSE) loss between the estimated noise $\hat{\boldsymbol{\epsilon}}_t$ and the true noise $\boldsymbol{\epsilon}_t$.

In the inference or sampling process, the model takes random noise concatenated with the MRI input image ($\mathbf{x}_{y}$)  and iteratively denoises the segmentation mask by estimating the noise $\hat{\boldsymbol{\epsilon}}_t$ at each timestep. 

During the sampling procedure, uncertainty maps are synthesized by exploiting the inherent stochasticity present in DDPMs. Through iterative application of IISMD, multiple segmentation masks are produced for a given input image. The uncertainty map is derived by assessing the pixel-wise variance of the  masks.

\begin{table}[h]
\caption{Training Hyperparameters for IISDM}
\centering
\small
\begin{tabular}{lccc}
\toprule
\textbf{Hyperparameter} & \textbf{T1w, T2w, T1w+T2w }  \\ \midrule
\textbf{Image Size} & 320x320 \\  
\textbf{Epochs} & 2600 \\  
\textbf{Batch} & 10  \\
\textbf{Optimizer} & AdamW  \\  
\textbf{Learning Rate} & 0.0001 \\ 
\textbf{Training Loss} & MSE \\ \bottomrule
\end{tabular}
\end{table}

\end{document}